\documentclass[nofootinbib]{revtex4}
\usepackage{graphicx}
\usepackage{latexsym}
\def\be{\begin{equation}}
\def\ee{\end{equation}}
\def\bea{\begin{eqnarray}}
\def\eea{\end{eqnarray}}

\begin{document}

\title{Avoiding the Big-Rip Jeopardy in a Quintom Dark Energy Model with Higher Derivatives}

\author{Xiao-fei Zhang\footnote{zhangxf@ihep.ac.cn} and Taotao Qiu\footnote{qiutt@ihep.ac.cn} }
 \affiliation{Institute of
High Energy Physics, Chinese Academy of Sciences, P.O. Box 918-4,
Beijing 100049, P. R. China}

\begin{abstract}

In the framework of a single scalar field quintom model with higher
derivative, we construct in this paper a dark energy model of which
the equation of state (EOS) $w$ crosses over the cosmological
constant boundary. Interestingly during the evolution of the
universe $w<-1$ happens just for a period of time with a
distinguished feature that $w$ starts with a value above $-1$,
transits into $w<-1$, then comes back to $w>-1$. This avoids the Big
Rip jeopardy induced by $w<-1$.

\end{abstract}

\maketitle

\hskip 1.6cm PACS number(s): {98.80.-k, 95.36.+x} \vskip 0.4cm

\section{Introduction}

  In 1998 the analysis of the redshift-distance relation of Type Ia
supernova (SNIa) established that our universe is currently
accelerating~\cite{Riess98,Perl99}, which has been further
confirmed by recent observations of SNIa at high confidence level
\cite{Tonry03,Riess04,Riess05}.  Combined with other observations
and experiments this strongly indicates that our universe is
dominated by a component with a negative pressure, dubbed dark
energy, which has been studied widely in the recent years. Various
candidates for dark energy, such as cosmological constant,
quintessence \cite{pquint,quint}, phantom \cite{phantom} and the
model of k-essence with non-canonical kinetic term
\cite{Chiba:1999ka,kessence}, have been proposed.

Though the recent analysis on the data from the Supernova, cosmic
microwave background (CMB) and large scale structure (LSS) show that
the cosmological constant fits well to the data, the dynamical
models of dark energy are generally not excluded, see Ref.
\cite{CST} for a recent review on dynamical dark energy and actually
a class of models, the quintom, with an EOS $w$ larger than $-1$ in
the past, less than $-1$ today, and evolving across $-1$ in the
intermediate redshift is mildly favored \cite{quintom, 26}.

There have been many increasing activities in the studies on the
quintom-like models of dark energy
\cite{guozk,hu,zhang,xfzhang,Wei:2005nw,a1,a2,a3,a4,a5,a6,a7,a8,a9,a10,a11,a12,a13,a14,a15,a16,a17,a18}
in the recent years. For the conventional single scalar field, the
EOS $w$ cannot cross over $w=-1$ \cite{Vikman,michael,Zhao:2005vj}
due to the instabilities of dark energy perturbation during
evolution. In this paper we focus on a particular example of the
quintom dark energy with a single scalar field and study its
cosmological evolution.

The model under investigation is a generalization of the model in
Ref.\cite{li} with the Lagrangian given by
\begin{equation}\label{lagrangian}
{\cal L}={\cal
L}(\phi,X,\Box\phi\Box\phi,\nabla_{\mu}\nabla_{\nu}\phi\nabla^{\mu}
\nabla^{\nu}\phi)~.
\end{equation}
In the lagrangian (\ref{lagrangian}),
$X\equiv\frac{1}{2}\nabla_{\mu}\phi \nabla^{\mu}\phi$ and
$\Box\equiv\nabla_{\mu}\nabla^{\mu}$. From (\ref{lagrangian}), it is
straightforward to get the equation of motion

\be \frac{\partial \mathcal{L}}{\partial
\phi}-\nabla_{\mu}(\frac{\partial \mathcal{L}}{\partial
X}\nabla^{\mu}\phi) +\Box(\frac{\partial\mathcal{L}}{\partial
\Box\phi})+\nabla_{\nu}\nabla_{\mu}(\frac{\partial\mathcal{L}}{\partial
s} \nabla^{\mu}\nabla^{\nu}\phi)=0~, \ee and the energy-momentum
tensor

\bea T^{\mu\nu}=& &[\nabla_{\rho}(\frac{\partial
\mathcal{L}}{\partial
\Box\phi}\nabla^{\rho}\phi)-\mathcal{L}]g^{\mu\nu} +\frac{\partial
\mathcal{L}}{\partial X}\nabla^{\mu}\phi\nabla^{\nu}\phi-
\nabla^{\mu}(\frac{\partial \mathcal{L}}{\partial
\Box\phi})\nabla^{\nu}\phi-
\nabla^{\nu}(\frac{\partial \mathcal{L}}{\partial \Box\phi})\nabla^{\mu}\phi- \nonumber\\
& & \nabla_{\rho}(\frac{\partial \mathcal{L}}{\partial
s}\nabla^{\mu}\nabla^{\rho}\phi)\nabla^{\nu}\phi-
\nabla_{\rho}(\frac{\partial \mathcal{L}}{\partial
s}\nabla^{\nu}\nabla^{\rho}\phi)\nabla^{\mu}\phi+
\nabla_{\rho}(\frac{\partial \mathcal{L}}{\partial
s}\nabla^{\mu}\nabla^{\nu}\phi)\nabla^{\rho}\phi~, \eea

where $s$ is defined as

\be s\equiv
\frac{1}{2}\nabla_{\mu}\nabla_{\nu}\phi\nabla^{\mu}\nabla^{\nu}\phi~.
\ee

This model for the moment is just an effective theory and we assume
that the operators associated with the higher derivatives can be
derived from some fundamental theories, for instance due to the
quantum corrections or the non-local physics in the string theory
\cite{16}\cite{17}\cite{18}. In addition, with the higher derivative
terms to the Einstein gravity, the theory is shown to be
renormalizable \cite{19} which has attracted many attentions.
Recently, higher derivative operators have been considered to
stabilize the linear fluctuations in the scenario of ``ghost
condensation" \cite{20}.

In Ref. \cite{li}, the authors studied in detail a specifical model
which is equivalent to the uncoupled two-field model (one is the
quintessence-like and the other is the phantom-like) and gives rise
to the fate of the universe dominated by the phantom-like field, as
was implied in two-field quintom model
\cite{quintom,guozk,hu,zhang,xfzhang,Wei:2005nw}. Thus the universe
might face some problems such as ending in ``big rip" \cite{big rip}
(There are some other possibilities of the fate of phantom universe,
see \cite{Sami,Shin'ichi,Abhik,sudden}). In this paper, however, we
generalize this kind of model and show some new possibilities of the
cosmological evolution. Specifically we will show  that in our case
there is a possibility that $w<-1$ happens just for a period of
time. Hence in this scenario it is free of the big rip jeopardy.

\section{Models with higher derivative}
Consider the Lagrangian with the following form
\begin{equation}\label{l1}
{\cal
L}=\frac{1}{2}A(\phi)\nabla_\mu\phi\nabla^\mu\phi+\frac{C(\phi)}{2M_{pl}^2}(\Box\phi)^2-V(\phi)~.
\end{equation}
As proved explicitly in \cite{li}, such a model is
classically equivalent to a two-field model. This is consistent with the general
result stated by Ostrogradsky theorem (a nice review on this theorem can be found in \cite{Ostrogradsky}),
higher derivative theory brings more degrees of freedom.
For the purpose of seeing how the model in (\ref{l1}) is equivalent to
two-field model, we introduce an auxiliary field $\chi$ and a
new Lagrangian, \be\label{l2}
\mathcal{L}'=\frac{1}{2}A(\phi)\nabla_\mu\phi\nabla^\mu\phi-\nabla_{\mu}\phi\nabla^{\mu}\chi-\frac{M_{pl}^2\chi^2}{2
C(\phi)} -V(\phi)~. \ee
Through the variation on $\chi$, one can see that
\be
\chi=\frac{C(\phi)}{M_{pl}^2}\Box\phi~. \ee By substituting it
into (\ref{l2}) and dropping out a total derivative,
 the new Lagrangian (\ref{l2}) reduces exactly to the old one
(\ref{l1}). But the former has the form of two-field without higher derivatives.
This is more clearly by further field transformations. Consider the field $\phi$ as a function of
two independent scalar fields $\chi$ and $\psi$. Using the
transformation \be\label{tran} \phi=\phi(\psi, \chi)~, \ee Eq.
(\ref{l2}) can be written as  \be\label{l3}
\mathcal{L}'=\frac{1}{2}A(\phi)(\frac{\partial \phi}{\partial
\psi})^2\nabla_{\mu}\psi\nabla^{\mu}\psi
-\frac{1}{2}A(\phi)(\frac{\partial \phi}{\partial
\chi})^2\nabla_{\mu}\chi\nabla^{\mu}\chi-\frac{M_{pl}^2\chi^2}{2
C(\phi)} -V[\phi(\psi, \chi)]~, \ee as long as the transformation
(\ref{tran}) satisfies \be\label{t1}
A(\phi)\frac{\partial\phi}{\partial\chi}=1~. \ee So the model in
(\ref{l1}) is equivalent to the two-field model (\ref{l3}). The
kinetic terms of $\psi$ and $\chi$ have to take opposite signs and
one of them must be ghost, which depends on the sign of $A(\phi)$.
Furthermore, we can choose $\phi$ as a function of $\psi+\chi$ and
set \be\label{t2} A(\phi)\frac{\partial\phi}{\partial\psi}=1~. \ee
So, we get a more elegant form for the Lagrangian \be\label{l4}
\mathcal{L}'=\frac{1}{2A[\phi(\psi+\chi)]}\nabla_{\mu}\psi\nabla^{\mu}\psi
-\frac{1}{2A[\phi(\psi+\chi)]}\nabla_{\mu}\chi\nabla^{\mu}\chi-\frac{M_{pl}^2\chi^2}{2
C[\phi(\psi+\chi)]} -V[\phi(\psi+\chi)]~. \ee

The form of $\phi$ as a function of $\chi$ and $\psi$ is determined
by the conditions (\ref{t1}) and (\ref{t2}), except constants.
However, there are some constraints on $A(\phi)$ and $C(\phi)$. As
indicated in (\ref{l2}), (\ref{t1}) and (\ref{t2}), $A(\phi)$ and
$C(\phi)$ should not be zero in the configuration space. Once the
functions $A(\phi)$, $C(\phi)$ and $V(\phi)$ and the initial
conditions are determined, the evolution of the system in (\ref{l1})
is fixed. Based on the discussions above, we can simplify the
analysis on the single field model (\ref{l1}) by analyzing the
two-field model (\ref{l4}).

\section{a specific case with temporary phantom phase}
 We will study the evolving behaviors of such type of models
for a simple case with  $A(\phi)=-1,~V(\phi)=0$. Thus the
Lagrangian is simplified as \be \label{13}
\mathcal{L}'=-\frac{1}{2}\nabla_{\mu}\psi\nabla^{\mu}\psi+
\frac{1}{2}\nabla_{\mu}\chi\nabla^{\mu}\chi-\frac{M_{pl}^2\chi^2}{2
C[\phi(\psi+\chi)]}~.\ee
From Eqs.(\ref{t1}) and (\ref{t2}), one
has the relation $\phi=-(\psi+\chi)$. The equations of motion for
the two fields are: \bea \label{eq1}
& &\Box \psi+\frac{M_{pl}^2C'(\psi+\chi)}{2C^2(\psi+\chi)}\chi^2=0\nonumber~,\\
& &\Box \chi-\frac{M_{pl}^2C'(\psi+\chi)}{2C^2(\psi+\chi)}\chi^2
+\frac{M_{pl}^2}{C(\psi+\chi)}\chi=0~, \eea  where the prime is
the derivative with respect to $\psi+\chi$. In the
Friedmann-Robertson-Walker universe, $\Box =\partial^2/\partial
t^2+3H\partial/\partial t$ with $H$ being the Hubble expansion
rate. Furthermore, one can easily get the density and the pressure
of dark energy, which are: \bea\label{eq2} \rho &=&
-\frac{\dot\psi^2}{2}+\frac{\dot\chi^2}{2}+
\frac{M_{pl}^2\chi^2}{2C(\psi+\chi)}~,\\p &=&
-\frac{\dot\psi^2}{2}+\frac{\dot\chi^2}{2}-\frac{M_{pl}^2\chi^2}{2C(\psi+\chi)}~,
\eea and the equation of state is: \bea\label{eq3} w ={p\over
\rho} =-1-\frac{\dot\psi^2-\dot\chi^2}
{-\frac{\dot\psi^2}{2}+\frac{\dot\chi^2}{2}+\frac{M_{pl}^2\chi^2}{2C(\psi+\chi)}}~,
\eea where the dot represents the derivative with respect to time.
 $\chi$ is quintessence-like and $\psi$ is
phantom-like. They couple to each other through the effective potential, i.e.,
the last term in the right hand of equation (\ref{13}):
\be
V_{eff}=\frac{M_{pl}^2\chi^2}{2
C[\phi(\psi+\chi)]}~.
\ee

The coupling function considered in this paper is \be
C[\phi(\psi+\chi)]=C_0[\frac{\pi}{2}+\arctan(\frac{-\lambda\phi}{M_{pl}})]
=C_0[\frac{\pi}{2}+\arctan(\frac{\lambda(\psi+\chi)}{M_{pl}})]~, \ee
thus\be
C'(\psi+\chi)=\frac{C_0\lambda}{M_{pl}[1+\frac{\lambda^2}{M_{pl}^2}(\psi+\chi)^2]}~.
\ee Because $C(\psi+\chi)$ is almost constant when
$\left|\psi+\chi\right|\gg 0$, $\psi$ and $\chi$ are nearly
decoupled at these regimes. We call them as ``weak coupling"
regimes. By contrast, the two fields couple tightly in the ``strong
coupling" regime where $\left|\psi+\chi\right|\sim 0$. In the weak
coupling regime, as shown in Eq. (\ref{eq1}), the phantom-like field
$\psi$ behaves as a massless scalar field and its energy density
$-(1/2)\dot\psi^2$ dilutes as $a^{-6}$, where $a$ is the scale
factor of the universe. The quintessence-like field $\chi$ has a
mass term with $m_{\chi}=M_{pl}/\sqrt{C(\psi+\chi)}$. Its behavior
is determined by the ratio of $m_{\chi}/H$. If $m_{\chi}\ll H$,
$\chi$ is slow-rolling and it behaves like a cosmological constant.
On the other hand in cases $m_{\chi}\gg H$, the kinetic term and
potential oscillate coherently and assembly evolve as $a^{-3}$, just
like that of collisionless dust.

It is however more complicated in the strong coupling regime. If the
effective potential $V_{eff}$ is negligible in comparison with the
energy density of the dominant component in the universe, as in the
period of radiation and matter domination, the Hubble damping terms
$3H\dot\psi$ and $3H\dot\chi$ dominates in the equations of motions,
see Eq. (\ref{eq1}), both $\psi$ and $\chi$ are ``frozen" and the
equation of state of dark energy $w\simeq -1$. If $V_{eff}$
contributes significantly to the energy density of  the universe
when dark energy is in the strong coupling regime, both of the
fields will relax to their equilibrium points with large velocities.
Because $C(\psi+\chi)$ and $C'(\psi+\chi)$ are positive, the
acceleration of $\psi$,
$\frac{M_{pl}^2C'(\psi+\chi)}{2C^2(\psi+\chi)}\chi^2$ is larger than
that of $\chi$:
$\frac{M_{pl}^2C'(\psi+\chi)}{2C^2(\psi+\chi)}\chi^2-\frac{M_{pl}^2}{C(\psi+\chi)}\chi$,
provided $\chi$ is positive. Hence, the ratio of
$\dot\psi^2/\dot\chi^2$ will increase and from Eq. (\ref{eq3}) the
equation of state $w$ will become less than $-1$ when
$\dot\psi^2>\dot\chi^2$. Otherwise, if $\chi$ is negative,
$\dot\psi^2/\dot\chi^2$ will decrease with the expansion of the
universe.

It is necessary to address more about the equilibrium points of
$\psi$ and $\chi$. Through simple algebraic calculations of
$\partial V_{eff}/\partial \psi=0$ and $\partial V_{eff}/\partial
\chi=0$, we find the equilibrium point of $\psi$ is in infinity
except for the line of $\chi= 0$. In the $\chi$ direction, the zero
point $\chi=0$ is the unique minimum. So, if the initial condition
is chosen as $\chi$ is not close to zero, the phantom-like field
$\psi$ will roll up the potential monotonically, this is along the
direction of $\psi\rightarrow -\infty$ and to reduce $C(\psi+\phi)$.
The quintessence-like field $\chi$ will relax down to the zero
point. In a word, the system will evolve to the weak coupling regime
and the field $\chi$ will get a large mass
$m_{\chi}=M_{pl}/\sqrt{C(\psi+\chi)}$ because
$C(\psi+\chi)\rightarrow 0$ as $\psi+\chi\rightarrow -\infty$. In
Fig.1, the effective potential (the green line), $\chi$ (the black
line), and $\psi$ (the red line) as functions of ln$a$ are plotted.

 We plot the EOS of such a dark energy model in Fig.2
and Fig.3 with different initial values. We choose such initial
conditions that the dark energy starts in the strong coupling regime
and $\chi>0$ at high red-shift. We can see that the dark energy is
frozen quickly at the initial time. Then it becomes significant
around the redshift of $z\sim 1$, and the phantom kinetic term
$\dot\psi^2$ becomes larger than the quintessence kinetic term
$\dot\chi^2$. The dark energy crosses the boundary of cosmological
constant and its equation of state becomes less than $-1$. In the
future, the system will evolve to the weak coupling regime,
$\dot\psi^2$ will dilute quickly and $\chi$ will get a large mass.
The whole system will behave as cold matter and its equation of
state will oscillate around the point of $w=0$. So, the universe
will exit from the phantom phase and end in the state of
matter-domination. This is consistent with the analysis above. Our
result resembles the late time behavior of ``B-inflation" model
\cite{Vikman2}.

\begin{figure}[htbp]
\includegraphics[scale=0.8]{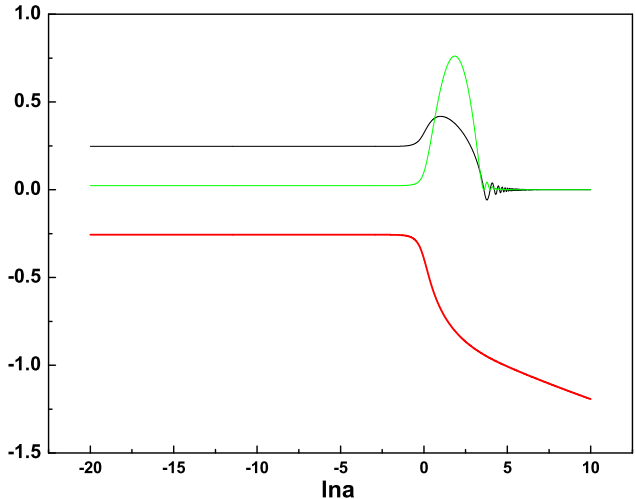}
\caption{the effective potential $V_{eff}\times C_{0}$ (the green
line), $\chi$ (the black line), and $\psi$ (the red line) as
functions of ln$a$ for
$C_0/M_{pl}^{2}=6.0\times10^{121}M_{pl}^{-2},~\lambda=25,$ and the
initial values are $
\psi_i=-0.26M_{pl},~\dot{\psi}_i=3.52\times10^{-62}M_{pl}^2,
~\chi_i=0.25M_{pl},~\dot{\chi}_i=-3.62\times10^{-62}M_{pl}^2,~with~\Omega_{DE_0}\approx0.73.$
}
\end{figure}

\begin{figure}[htbp]
\includegraphics[scale=0.8]{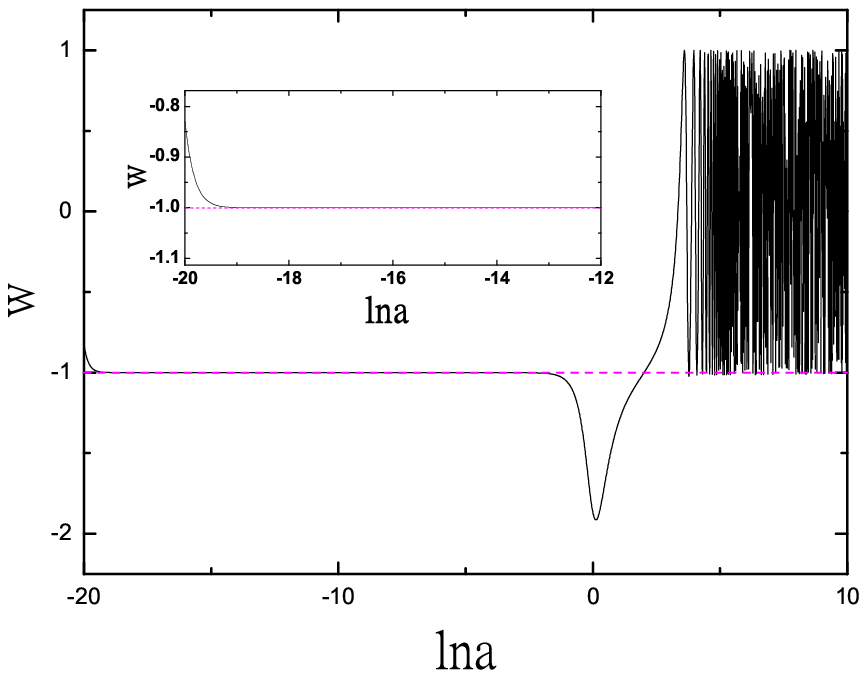}
\caption{The equation of state $w$ as a function of ln$a$ for
$C/M_{pl}^{2}=6.0\times10^{121}M_{pl}^{-2},~\lambda=25,$ and the
initial values are $
\psi_i=-0.26M_{pl},~\dot{\psi}_i=3.52\times10^{-62}M_{pl}^2,
~\chi_i=0.25M_{pl},~\dot{\chi}_i=-3.62\times10^{-62}M_{pl}^2,~
with~\Omega_{DE_0}\approx0.73.$ }
\end{figure}

\begin{figure}[htbp]
\includegraphics[scale=0.8]{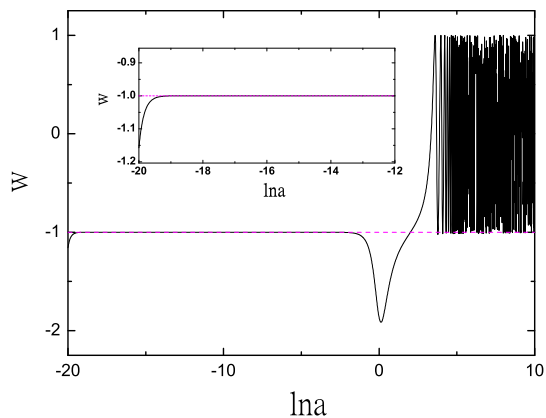}
\caption{The equation of state $w$ as a function of ln$a$ for
$C/M_{pl}^{2}=6.0\times10^{121}M_{pl}^{-2},~\lambda=25,$ and the
initial values are $
\psi_i=-0.26M_{pl},~\dot{\psi}_i=-2.84\times10^{-62}M_{pl}^2,
~\chi_i=0.25M_{pl},~\dot{\chi}_i=2.74\times10^{-62}M_{pl}^2
,~with~\Omega_{DE_0}\approx0.73.$ }
\end{figure}

The cosmological
meanings of such a model are as follows:

$\bullet$ In this model, we give a kind of cosmological scenario
which differs from the former quintom dark energy models. This
model can give twice crossing of $w$, and can give a nice exit of
phantom which stays just for a period of time, avoiding ``big rip"
to happen.

$\bullet$ This kind of $w$ gives an example of feature EOS in
model building in field theory, which is being paid more and more
attention on by current data-fitting \cite{feature}.

$\bullet$ At late time, $w$ oscillates with high frequency around
zero, so the universe will end accelerated expanding at some time.

\section{summary}

In summary, we pointed out that for quintom model of single scalar
field with higher derivative, the EOS $w$ can have some interesting
behaviors. We provide a scenario where the EOS $w$ starts from above
$-1$, transits to below $-1$, then comes back to above $-1$. This in
some sense implies that $w<-1$ may happen just for a period of time.
With this behavior of $w$, this scenario might be able to get rid of
the problem caused by phantom. For phantom theory, the EOS $w$ is
always less than $-1$, and there might be many problems such as
ending in ``big rip" as well as other singularities, and quantum
instability. However in some class scenario of quintom with higher
derivative, the evolution of $w<-1$ is very much like the ``tachyon"
existing only during the phase transition \cite{x.m.zhang talk}.
Thus the ``big rip" problem \cite{big rip} is solved in such a
context.

The quantum instability of phantom universe is discussed in detail
in Ref. \cite{James,Sean}. Our quintom model based on higher
derivative theory also has ghost degree and the problem of quantum
instability. However, as pointed out in Ref. \cite{hawking}, the
problem of quantum instability arises because $\phi$ and $\Box\phi$
are quantized in canonical way independently. In fact, both of them
are determined by $\phi$, and a more appropriate quantization method
seems to be possible to avoid the instability.

This work not only provides some new
examples for the list of field theory model of dark energy, which
deserves being included in a more thorough classification, but
useful for stimulating the studies on the field theory with higher
derivatives.

\section{acknowledgements}
We thank Prof. Xinmin Zhang for patient guidance and Bo Feng, Hong Li, Mingzhe Li, Yun-Song Piao,
Jun-Qing Xia and Gong-Bo Zhao for
helpful discussions. This work is supported in part by National
Natural Science Foundation of China under Grant Nos. 90303004,
10533010 and 19925523 and by Ministry of Science and Technology of
China under Grant No. NKBRSF G19990754.


\begin{thebibliography}{99}
\expandafter\ifx\csname
natexlab\endcsname\relax\def\natexlab#1{#1}\fi
\expandafter\ifx\csname bibnamefont\endcsname\relax
  \def\bibnamefont#1{#1}\fi
\expandafter\ifx\csname bibfnamefont\endcsname\relax
  \def\bibfnamefont#1{#1}\fi
\expandafter\ifx\csname citenamefont\endcsname\relax
  \def\citenamefont#1{#1}\fi
\expandafter\ifx\csname url\endcsname\relax
  \def\url#1{\texttt{#1}}\fi
\expandafter\ifx\csname urlprefix\endcsname\relax\def\urlprefix{URL
}\fi \providecommand{\bibinfo}[2]{#2}
\providecommand{\eprint}[2][]{\url{#2}}


\bibitem{Riess98}
A.G. Riess {\it et al.} (Supernova Search Team Collaboration),
Astron. J. {\bf 116}, 1009 (1998).

\bibitem{Perl99}
S. Perlmutter {\it et al.} (Supernova Cosmology Project
Collaboration), Astrophys. J. {\bf 517}, 565 (1999).

\bibitem{Tonry03}
J. L. Tonry {\it et al.} (Supernova Search Team Collaboration),
Astrophys. J. {\bf 594}, 1 (2003).

\bibitem{Riess04}
A.~G.~Riess {\it et al.}  (Supernova Search Team Collaboration),
Astrophys.\ J.\  {\bf 607}, 665 (2004).
% astro-ph/0402512.

\bibitem{Riess05}
A.~Clocchiatti {\it et al.}  (the High Z SN Search Collaboration),
astro-ph/0510155.



\bibitem{pquint}
B. Ratra and P. J. E. Peebles, Phys. Rev. D {\bf 37}, 3406 (1988);
P. J. E. Peebles and B. Ratra, Astrophys. J. {\bf 325}, L17 (1988).

\bibitem{quint}
R.~D.~Peccei, J.~Sola and C.~Wetterich,  Phys.\ Lett.\ B {\bf 195},
183 (1987); C. Wetterich, Nucl. Phys. B {\bf 302}, 668 (1988); C.
Wetterich, Astron. Astrophys. {\bf 301}, 321 (1995).


\bibitem{phantom}
R. R. Caldwell, Phys. Lett. B {\bf 545}, 23 (2002).



\bibitem{Chiba:1999ka}
  T.~Chiba, T.~Okabe and M.~Yamaguchi,

  Phys.\ Rev.\ D {\bf 62} (2000) 023511.


\bibitem{kessence}
C. Armendariz-Picon, V. Mukhanov and P. J. Steinhardt, Phys. Rev.
Lett. {\bf 85}, 4438 (2000); Phys. Rev. D {\bf 63}, 103510 (2001).


\bibitem{CST} E. J. Copeland, M. Sami and S. Tsujikawa,
hep-th/0603057.

\bibitem{quintom} B. Feng, X. Wang and X. Zhang, Phys. Lett. B {\bf 607},
35, (2005).

\bibitem{26}
For recent studies see e.g.  G. B. Zhao, J. Q. Xia, B. Feng and X.
Zhang, astro-ph/0603621; X. Zhang and F.-Q. Wu, Phys. Rev. D {\bf
72}, 043524 (2005); Z.Chang, F.-Q. Wu and X. Zhang, Phys. Lett. B
{\bf 633}, 14 (2006).

\bibitem{guozk}
Z. K. Guo, Y. S. Piao, X. Zhang and Y. Z. Zhang, Lett. B {\bf
608}, 177, (2005).
\bibitem{hu} W. Hu, Phys. Rev. D71 (2005) 047301. %astro-ph/0410680.

\bibitem{zhang} X. Zhang, hep-ph/0410292.

\bibitem{xfzhang} X. F. Zhang, H. Li, Y. S. Piao, and X. Zhang,
Mod. Phys. Lett. A {\bf 21}, 231 (2006).  %astro-ph/0501652.
\bibitem{Wei:2005nw}
  For an interesting variation see H.~Wei, R.~G.~Cai and D.~F.~Zeng,
Class.\ Quant.\ Grav.\  {\bf 22}, 3189 (2005).

\bibitem{a1}
 Diego F. Torres, Phys.Rev.D {\bf 66}, 043522 (2002).

\bibitem{a2}
 Rong-Gen Cai, Anzhong Wang,
 JCAP {\bf 0503}, 002 (2005).

\bibitem{a3}
 L. Perivolaropoulos, Phys.Rev.D {\bf 71}, 063503 (2005).

\bibitem{a4}
Shin'ichi Nojiri, Sergei D. Odintsov, and Shinji Tsujikawa,
Phys.Rev.D {\bf 71}, 063004 (2005).

\bibitem{a5}
Hrvoje Stefancic, Phys.Rev.D {\bf 71}, 124036 (2005).

\bibitem{a6}
L. Perivolaropoulos, JCAP {\bf 0510}, 001 (2005).

\bibitem{a7}
Alexander A. Andrianov, Francesco Cannata, and Alexander Y.
Kamenshchik, Phys.Rev.D {\bf 72}, 043531 (2005).

\bibitem{a8}
I.Ya. Aref'eva, A.S. Koshelev, and S.Yu. Vernov, Phys.Rev.D {\bf
72}, 064017 (2005); Chao-Guang Huang, Han-Ying Guo,
astro-ph/0508171.





\bibitem{a9}
Hrvoje Stefancic, J.Phys.A {\bf 39}, 6761-6768 (2006).

\bibitem{a10}
S. Nesseris, L. Perivolaropoulos, Phys.Rev.D {\bf73}, 103511
(2006).

\bibitem{a11}
Ruth Lazkoz, Genly Le¨®n, astro-ph/0602590.

\bibitem{a12}
Pantelis S. Apostolopoulos, Nikolaos Tetradis, hep-th/0604014.

\bibitem{a13}
Luis P. Chimento, Ruth Lazkoz, astro-ph/0604090.

\bibitem{a14}
Wen Zhao, Yang Zhang, Phys.Rev.D {\bf 73}, 123509 (2006).

\bibitem{a15}
Bin Wang, Yungui Gong, and Elcio Abdalla, Phys.Lett.B {\bf 624},
141-146 (2005).

\bibitem{a16}
Shinji Tsujikawa, Phys.Rev.D {\bf 72}, 083512 (2005).

\bibitem{a17}
Wayne Hu, Phys.Rev. D {\bf 71}, 047301 (2005).

\bibitem{a18}
Robert R. Caldwell, Michael Doran, Phys.Rev. D {\bf 72}, 043527
(2005).

\bibitem{Vikman}
A. Vikman, Phys.\ Rev.\ D {\bf 71}, 023515 (2005). %astro-ph/0407107.

\bibitem{michael} R. R. Caldwell and M. Doran, Phys.\ Rev.\ D {\bf 72}, 043527
(2005).
%astro-ph/0501104.

\bibitem{Zhao:2005vj}
  G.~B.~Zhao, J.~Q.~Xia, M.~Li, B.~Feng and X.~Zhang,
  Phys. Rev. D {\bf 72}, 123515 (2005).
%astro-ph/0507482.
  %%CITATION = ASTRO-PH 0507482;%%


\bibitem{li}
M. Li, B. Feng and X. Zhang, JCAP {\bf 0512}, 002 (2005).% [hep-ph/0503268].

\bibitem{16}
J.Z. Simon, Phys. Rev. D {\bf 41}, 3720 (1990).

\bibitem{17}
A. Elizer and R. P. Woodard, Nucl. Phys. B {\bf 325}, 389 (1989).

\bibitem{18}
T. G. Erler and D. J. Gross, hep-th/0406199.

\bibitem{19}
K. S. Stelle, Phys. Rev. D {\bf 16}, 953 (1977).

\bibitem{20}
 N. Arkani-Hamed, H. C. Cheng, M. A. Luty and S. Mukohyama, JHEP {\bf 0405}, 074 (2004); % [hep-th/0312099];
 N. Arkani-Hamed, P. Creminelli, S. Mukohyama and M. Zaldarriaga, JCAP {\bf 0404}, 001 (2004);
 Federico Piazza and Shinji Tsujikawa, hep-th/0405054;
 A. Anisimov and A. Vikman, JCAP {\bf 0504}, 009 (2005); % [hep-ph/0411089];
 S. Mukohyama, Phys. Rev. D {\bf 71}, 104019 (2005); % [hep-th/0502189];
 N. Arkani-Hamed, H. C. Cheng, M. A. Luty, S. Mukohyama and T. Wiseman,
 hep-ph/0507120.



\bibitem{big rip}  R. R. Caldwell, M. Kamionkowski, and N. N.
Weinberg,
Phys. Rev. Lett. {\bf 91}, 071301  (2003). %astro-ph/0302506
\bibitem{Sami}
M. Sami and Alexey Toporensky, Mod.Phys.Lett. A {\bf19}, 1509
(2004).

\bibitem{Shin'ichi}
Shin'ichi Nojiri and Sergei D. Odintsov, Phys.Lett. B{\bf595}  1-8
(2004).

\bibitem{Abhik}
Abhik Kumar Sanyal, astro-ph/0605388.

\bibitem{sudden}
Mariusz P. Dabrowski, Annalen Phys{\bf15}, 352-363 (2006),
astro-ph/0606574.

%\bibitem{wmap new} D. N. Spergel, et al, astro-ph/0603449.
\bibitem{Ostrogradsky}
R. P. Woodard, astro-ph/0601672.

\bibitem{Vikman2}
A.~Anisimov, E.~Babichev and A.~Vikman,
  %``B-inflation,''
  JCAP {\bf 0506}, 006 (2005).

\bibitem{feature} J.~Q.~Xia, G.~B.~Zhao, H.~Li, B.~Feng and X.~Zhang, astro-ph/0605366.
\bibitem{x.m.zhang talk} X. Zhang, AIP Conf.Proc.805:3-9,2006;
 Also in *Gyeongju 2005, Particles, strings and cosmology* 3-9,
 hep-ph/0510072.

\bibitem{James}
James M. Cline, Sangyong Jeon and Guy D. Moore, Phys. Rev. D
{\bf70},  043543 (2004).

\bibitem{Sean}
Sean M. Carroll, Mark Hoffman and Mark Trodden, Phys.Rev. D
{\bf68}, 023509 (2003).

\bibitem{hawking}
S. W. Hawking and  T. Hertog, Phys. Rev. D  {\bf 65}, 103515 (2002). % [arXiv:hep-th/0501160].

\end{thebibliography}
\end{document}